\documentstyle[aps,prd,floats,epsf]{revtex}
\tighten
\draft
\begin{document}

\twocolumn[\hsize\textwidth\columnwidth\hsize\csname
@twocolumnfalse\endcsname
\title{Can a supernova be located by its neutrinos?}
\author{J.~F. Beacom\thanks{Electronic address:
        {\tt beacom@citnp.caltech.edu}} and
        P. Vogel\thanks{Electronic address:
        {\tt vogel@lamppost.caltech.edu}}}
\address{Department of Physics, California Institute of Technology\\
         Pasadena, CA 91125, USA}
\date{November 21, 1998; revised version May 17, 1999}
\maketitle

\begin{abstract}

A future core-collapse supernova in our Galaxy will be detected by several
neutrino detectors around the world.  The neutrinos escape from the
supernova core over several seconds from the time of collapse, unlike
the electromagnetic radiation, emitted from the envelope, which is
delayed by a time of order hours.  In addition, the electromagnetic
radiation can be obscured by dust in the intervening interstellar
space.  The question therefore arises whether a supernova can be
located by its neutrinos alone.  The early warning of a supernova and
its location might allow greatly improved astronomical observations.
The theme of the present work is a careful and realistic assessment of
this question, taking into account the statistical significance of the
various neutrino signals.  Not surprisingly, neutrino-electron forward
scattering leads to a good determination of the supernova direction,
even in the presence of the large and nearly isotropic background from
other reactions.  Even with the most pessimistic background
assumptions, SuperKamiokande (SK) and the Sudbury Neutrino Observatory
(SNO) can restrict the supernova direction to be within circles of
radius $5^\circ$ and $20^\circ$, respectively.  Other reactions with
more events but weaker angular dependence are much less useful for
locating the supernova.  Finally, there is the oft-discussed
possibility of triangulation, i.e., determination of the supernova
direction based on an arrival time delay between different detectors.
Given the expected statistics we show that, contrary to previous
estimates, this technique does not allow a good determination of the
supernova direction.

\end{abstract}

\pacs{97.60.Bw, 95.55.Vj, 13.10.+q, 25.30.Pt}

\vspace{0.2cm}]
\narrowtext


\section{Introduction}

There has been great interest recently in the question of whether or
not a supernova can be located by its neutrinos.  If so, this may
offer an opportunity to give an early warning to the astronomical
community, so that the supernova light curves can be observed from the
earliest possible time.  An international supernova early alert
network has been formed for this purpose, and the details of its
implementation~\cite{Habig,nu98} were the subject of a recent
workshop~\cite{SNwksp}.  (The creation of such a network was
also discussed in Ref.~\cite{Cline}).
One of the primary motivations for such a network is to greatly reduce
the false signal rate by demanding a coincidence between several
different detectors, as detailed in Refs.~\cite{Habig,nu98}.  The
second motivation, to locate the supernova by the neutrino signal, 
is the topic of this paper.

An early-warning network is important because supernovae are rare,
with the estimated core-collapse supernova rate in the Galaxy about 3
times per century~\cite{SNrate}.  The present neutrino detectors can
easily observe a supernova anywhere in the Galaxy or its immediate
companions (e.g., the Magellanic Clouds).  Unfortunately, the present
detectors do not have large enough volumes to observe a supernova in
even the nearest galaxy (Andromeda, about 700 kpc away).

Since decisions about how to implement this network are being made
now, it is of current and necessary practical interest to make
detailed calculations of what can realistically be done.  In this
paper we carefully examine the available techniques for locating the
supernova by its neutrino signal.  The problem has been discussed in
general before (see Ref.~\cite{Burrows}, for example).  (See also the
early discussions in Refs.~\cite{Zatsepin1,Zatsepin2,LoSecco}).
Previous estimates, in particular regarding the triangulation
problem~\cite{Habig,nu98,Burrows,Vagins,Virtue}, were rather optimistic.
In this paper, we make explicit some of the underlying assumptions in
these calculations, and explain what the fundamental limitations on
the precision are.  Generally speaking, we find that this problem is
more difficult than had been anticipated.  The results below were
first presented at the above workshop~\cite{Beacom}.

There are two types of techniques to locate a supernova by its
neutrinos.  The first class of techniques is based on angular
distributions of the neutrino reaction products, which can be
correlated with the neutrino direction.  In this case, a single
experiment can independently announce a direction and its error.
The second method of supernova
location is based on triangulation using two or more widely-separated
detectors.  This technique would require significant and immediate
data sharing among the different experiments.  Our conclusion, that
triangulation is at best very crude, has an important impact on the
ongoing design decisions for the type of alert network and the degree
of data sharing.

Let us note in passing that the supernova might be not only located on
the sky, i.e., in two dimensions, but that its distance can be also
reasonably estimated.  The number of neutrino events $N$
is proportional to the binding energy release $E_B$
of the supernova, and of course falls off as the distance $D$ squared:
\begin{equation}
N = N_0 
\left(\frac{E_B}{3 \times 10^{53} {\rm\ ergs}}\right)
\left(\frac{10 {\rm\ kpc}}{D}\right)^2\,,
\end{equation}
where $N_0$ is the number of events at the canonical values of $E_B$
and $D$.  The binding energy is thought to be $E_B = (3.0 \pm 1.5)
\times 10^{53}$ ergs \cite{SNbinding}, i.e., a relative precision of
50\%.  The total numbers of events from all reactions are $N_0 \simeq
10^4$ for SK and $N_0 \simeq 10^3$ for SNO (the expected signals in SK
and SNO are discussed in Refs.~\cite{SKpaper,SNOpaper}).  Note that
$N_0$ depends on the neutrino spectrum temperatures (which can be
measured).  The relative errors on the measured $N$ and the calculated
$N_0$ are much smaller than the relative error on $E_B$.  Then
\begin{equation}
\frac{\delta D}{D} \simeq \frac{1}{2} \frac{\delta E_B}{E_B}
\end{equation}
and thus $D$ can be determined with a relative precision of order 25\%
by any detector with reasonable statistics.

Finally, it is likely that the next supernova will lie in the Galactic
plane (including the bulge).  However, in our opinion, the point of
the neutrino measurement is to make an unbiased estimate of the
supernova location, so we do not use this as a constraint.  For
concreteness, we assume that $D = 10$ kpc (approximately the distance
to the Galactic center), but in an arbitrary direction.


\section{Reactions with angular dependence}

Neutrinos are detected by their interaction with the target material,
i.e., its electrons or nuclei.  For some reactions, the angular
distribution of the reaction products is correlated with the incoming
neutrino direction.  In this section we describe how this angular
dependence can be used for determination of the supernova direction.


\subsection{Neutrino-electron scattering: forward peaking}

Neutrino-electron scattering,
\begin{equation}
\nu + e^- \rightarrow \nu + e^- \,,
\end{equation}
occurs for all flavors of neutrinos and antineutrinos, and is detected
by observing the recoil electrons with kinetic energy $T$ above the
experimental threshold $T_{\rm min}$.  The scattering angle is
dictated by the kinematics and is given by
\begin{equation}
\cos\alpha = \frac{E_{\nu} + m_e}{E_{\nu}} 
\left( \frac{ T}{T + 2m_e} \right)^{1/2}\,.
\end{equation}
Both SK and SNO hope to have a threshold of order $T_{\rm min} = 5$
MeV, and so $\cos\alpha \gtrsim 0.91$.  However, after integrating
over the electron kinetic energy distribution for a fixed neutrino
energy, and also the neutrino energy spectrum, the average value will
be larger.  We take the latter to be of the Fermi-Dirac type with
temperature $T = $ 3.5, 5 and 8 MeV for $\nu_e, \bar{\nu}_e$ and
$\nu_x \equiv \nu_{\mu}, \bar{\nu}_{\mu}, \nu_{\tau},
\bar{\nu}_{\tau}$, respectively.  Then we obtain $\langle \cos\alpha
\rangle$ = 0.98, 0.97, and 0.98, respectively, with the combined
average $\langle \cos \alpha \rangle$ = 0.98, corresponding to about
$11^{\circ}$.

The angular distribution of the produced electrons is narrow, and
depends on energy and flavor.  However, multiple scattering of the
electron will smear its \v{C}erenkov cone.  This washes out the
dependence on energy and flavor, and one can reasonably model the
electrons as having a Gaussian smearing from the forward direction,
with a one-sigma width of $25^\circ$, for all energies and flavors.
This is consistent with the estimate for SNO~\cite{SNOangular} and the
measurement by a LINAC for SK~\cite{SKangular}.  The SK measurement
shows that the angular resolution does depend on the electron energy,
but the variation is not large.

Naively, if the one-sigma angular width of this cone is $\delta \alpha
\simeq 25^\circ$, then the precision with which its center (i.e., the
average) can be defined given $N_S$ events is
\begin{equation}
\delta \theta \simeq \frac{\delta\alpha}{\sqrt{N_S}}\,,
\end{equation}
where $\theta$ measures the angle from the best-fit direction (i.e.,
the average).  For SK~\cite{SKpaper}, $N_S \simeq 320$, so the cone
center could be defined to within about $1.5^\circ$.  For SNO (using
both the light and heavy water)~\cite{SNOpaper}, $N_S \simeq 25$, so
the cone center could be defined to about $5^\circ$.  These results
(at least for SK) are widely known (see, for example,
Ref.~\cite{Burrows}).  The equivalent error on the cosine is
$\delta(\cos\theta) \simeq (\delta\theta)^2/2$, i.e., $3 \times
10^{-4}$ and $4 \times 10^{-3}$, respectively.

These results neglect the fact that the centroiding is to be done in
two dimensions, and that this peak sits on a large background.  It has
been claimed that the error on centroiding in two dimensions is larger
than the corresponding error in one dimension by a factor of
$\sqrt{2}$.  For $r = \sqrt{x^2 + y^2}$, and uncorrelated errors of
equal magnitude, $\delta x = \delta y = \sigma$, simple error
propagation gives $\delta r = \sigma$.  Only for correlated errors,
e.g., a positive error $\delta x$ always accompanied by a positive
$\delta y$ of the same magnitude, does the factor $\sqrt{2}$ appear.
Centroiding in two dimensions is no harder than centroiding in one
dimension since there are twice as many measurements, i.e., $x$ and
$y$ for each point.

However, the nearly isotropic background from all other reactions,
neglected in previous estimates, is more of a concern.  Finding the
supernova direction becomes a question of finding the centroid of a
Gaussian peak of known width on a known flat background.  The
centroiding precision can be determined by a test due to
DuMond~\cite{DuMond}.  This result follows from the assumption that a
known template function with unknown centroid is adjusted until it
gives the best least-squares fit to the data, i.e., exactly what one
would do in practice.  This technique does not require any arbitrary
cuts on which data are included in determining the centroid (which
would introduce bias).  The appropriate two-dimensional generalization
is
\begin{equation}
\frac{1}{(\delta x)^2} = \int dy \int dx \,
\frac{\left[\partial L(x,y)/\partial x\right]^2}{L(x,y)}\,,
\label{eq:error}
\end{equation}
where $L(x,y) = d^2 N/dx dy$, the density of events.  Note that since
the error is the same in any direction, and the $x$ and $y$ directions
are arbitrary, we have considered the case in which the centroiding
error $\delta x$ is only in the $x$ direction.  One can show that
this $\delta x$ is exactly the shift in the centroid which would 
increase the total $\chi^2$ of the fit by 1.  (One can also use the
Rao-Cramer theorem, introduced in a later section, to show that this
error $\delta x$ is the minimum which can be achieved by any
technique).

We use a two-dimensional Gaussian peak (with a total of $N_S$ signal
events) on a flat background:
\begin{equation}
L(x,y) = \frac{N_S}{2\pi\sigma^2}
\left[\exp\left(-\frac{x^2}{2 \sigma^2}\right)
\exp\left(-\frac{y^2}{2 \sigma^2}\right) + R\right]\,,
\end{equation}
where $R$ is the ratio of the heights of the flat background and the
signal (at peak).  For $N_B$ background events on the whole sphere,
\begin{equation}
R = \frac{\sigma^2}{2} \frac{N_B}{N_S}\,,
\end{equation}
where $\sigma = \delta\alpha \simeq 25^\circ$.  Once $R$ has been
specified, we treat the problem in a plane instead of on the sphere,
since $\sigma$ is small.  When $R = 0$, the integral can be done
analytically, with the expected result of $\delta x =
\sigma/\sqrt{N_S}$.  For $R > 0$, it must be done numerically.  We
define a correction factor $C(R)$ to the naive error by
\begin{equation}
C(R) = \frac{\delta x}{\sigma/\sqrt{N_S}}\,,
\end{equation}
where the full error $\delta x$ is determined numerically from
Eq.~(\ref{eq:error}).  The function $C(R)$ is shown in Fig.~1.  Note
that $C(R)$ depends on $R$ alone, so that the same figure can be used
for both SK and SNO.  Empirically, $C(R)$ is reasonably fit by $C(R)
\simeq \sqrt{1 + 4 R}$.  This form is motivated by two constraints:
that $C(0) = 1$, and that for $R \gg 1$, one can show from
Eq.~(\ref{eq:error}) that $C(R) \simeq \sqrt{4 R}$.

For SK and SNO, the number of signal events $N_S$ is 320 and 25,
respectively.  As a worst case, we assume that all events from other
reactions are background events.  Then for SK, $R \simeq 3.0$ and
$C(R) \simeq 3.7$, and for SNO, $R \simeq 3.8$ and $C(R) \simeq 4.1$.
These correction factors may seem surprisingly large, but one should
note that a two-sigma region contains about 2000 and 200 background
events for SK and SNO, respectively.  With the most pessimistic
background assumptions, the centroiding errors for SK and SNO are then
about $5^\circ$ and $20^\circ$, respectively.

It should be possible to reduce this isotropic background.  In
neutrino-electron scattering, the outgoing electrons tend to have
energies well below the neutrino energy.  In contrast, in the reaction
$\bar{\nu}_e + p \rightarrow e^+ + n$, the outgoing positron carries
almost all of the neutrino energy.  Approximately 2/3 of these
background events are above 20 MeV, and can be cut with little loss in
signal.  The background in the SNO heavy water will depend on the
neutron detection technique.  Crude estimates indicate that SK and SNO
may each be able to attain $C(R) \simeq 2 - 3$.

\begin{figure}[t]
\epsfxsize=3.25in \epsfbox{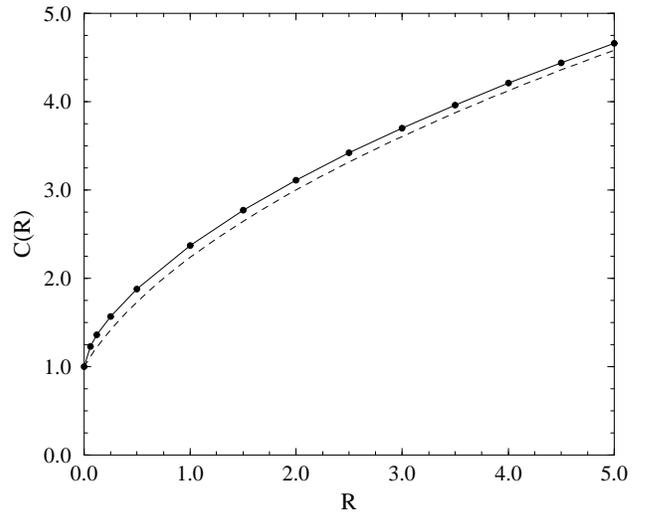}
\caption{The correction factor to the naive pointing error from $\nu +
e^- \rightarrow \nu + e^-$ scattering due to the isotropic background
is shown.  The solid line with points is the numerical result.  The
dashed line is $\protect\sqrt{1 + 4 R}$, an approximation that is
discussed in the text.}
\end{figure}


\subsection{Neutrino-nucleus reactions: weak angular dependence}

In this subsection we consider charged-current reactions on nuclear
targets, i.e., reactions in which only $\nu_e$ and $\bar{\nu}_e$
participate.  The reaction with the most events is $\bar{\nu}_e + p
\rightarrow e^+ + n$, with $\simeq 10^4$ events expected in SK, and
$\simeq 400$ events expected in the light water of SNO.  The other
relevant reactions are those on deuterons in SNO, with $\simeq$ 80
events each expected for $\bar{\nu}_e + d \rightarrow e^+ + n +n $ and
$\nu_e + d \rightarrow e^- + p +p$ \cite{SNOpaper}.  We neglect
charged-current reactions on the isotopes of oxygen~\cite{Haxton},
which also have weak asymmetries, but are difficult to separate from
more dominant reactions.  The reactions considered have lepton angular
distributions approximately of the form
\begin{equation}
\frac{d N}{d \cos\alpha} = \frac{N}{2} \left(1 + a \cos\alpha \right)\,,
\label{eq:acos}
\end{equation}
with, in general, $a = a(E_\nu)$.  We will discuss the magnitude and
variation of the coefficient $a$ shortly.  We first consider how well
one could localize the supernova assuming that $a$ is known and
constant.  In the following, we neglect the experimental angular
resolution for electron and positron directions, as it will be
negligible in comparison to the pointing errors discussed below.

Given a sample of events, one can attempt to find the axis defined by
the neutrino direction.  Along this axis, the distribution should be
flat in the azimuthal angle $\phi$ and should have the form of
Eq.~(\ref{eq:acos}) in $\cos\alpha$.  Along any other axis, the
distribution will be a complicated function of both the altitude and
azimuthal angles.  We assume that the axis has been found numerically,
and ask how well the statistics allow the axis to be defined.  A
convenient way to assess that is to define the forward-backward
asymmetry as
\begin{equation}
A_{FB} = \frac{N_F - N_B}{N_F + N_B}\,,
\end{equation}
where $N_F$ and $N_B$ are the numbers of events in the forward and
backward hemispheres.  The total number of events is $N = N_F + N_B$.
Note that $A_{FB}$ will assume an extremal value $A_{FB}^{extr}$
along the correct neutrino direction.

The error on $A_{FB}$ due to the statistical errors on $N_F$ and $N_B$ is
\begin{equation}
\delta A_{FB} = 
\frac{1 - A^2_{FB}}{2} \sqrt{\frac{1}{N_F} + \frac{1}{N_B}}\,.
\end{equation}
Using Eq.~(\ref{eq:acos}) one finds simply,
\begin{equation}
N_{F,B} = \frac{N}{2} \left(1 \pm \frac{a}{2}\right)\,.
\end{equation}
Therefore
\begin{equation}
A_{FB} = \frac{a}{2}
\end{equation}
and
\begin{equation}
\delta A_{FB}
= \frac{1}{\sqrt{N}} \sqrt{1 - \left(\frac{a}{2}\right)^2}
\simeq \frac{1}{\sqrt{N}}\,,
\end{equation}
where the error is nearly independent of $a$ for small $|a|$, which is
the case for the reactions under consideration.

In the above, the coordinate system axis was considered to be
correctly aligned with the neutrino direction.  Now consider what
would happen if the coordinate system were misaligned.  While in
general, all three Euler angles would be needed to specify an
arbitrary change in the coordinate system, symmetry considerations
dictate that the computed value of $A_{FB}$ depends only upon one --
the angle $\theta$ between the true and the supposed neutrino axis.
Thus $A_{FB}$ is some function of $\theta$ if the axis is misaligned.
Using a Legendre expansion, one can show that
\begin{equation}
A_{FB}(\theta) = \frac{a}{2} \cos\theta\,.
\label{eq:Afb}
\end{equation}
The same expression is obtained by expressing the angular distribution
in spherical harmonics and considering its behavior when acted on by
the rotation operator.  The error on the alignment is then
\begin{equation}
\delta (\cos\theta) =
\frac{2}{|a|} \, \delta A_{FB} \simeq
\frac{2}{|a|} \frac{1}{\sqrt{N}}\,.
\end{equation}

As noted, one would find the best estimate of the neutrino axis
numerically and define that direction to be $\cos\theta = 1$.  Along
that axis, the measured asymmetry will be $A_{FB}^{extr}$.  The above
considerations describe the situation when the fluctuations do not
dominate the value of $A_{FB}^{extr} \simeq a/2 \pm 1/\sqrt{N}$.  Only
in that case one can hope to use the angular distribution for
pointing, and at the same time avoid apparent formal difficulties like
an infinite error in the cosine when $a \rightarrow 0$ or a
possibility of $|\cos\theta| > 1$ in Eq.~(\ref{eq:Afb}).

Treating the nucleons as infinitely heavy, the coefficient $a$ in Eq.
(\ref{eq:acos}) is related to the competition of the Fermi (no spin
flip) and Gamow-Teller (spin flip) parts of the matrix element
squared:
\begin{equation}
a = \frac{ |M_F|^2 - |M_{GT}|^2}{ |M_F|^2 + 3 |M_{GT}|^2} \,.
\end{equation}
For the $\bar{\nu}_e + p \rightarrow e^+ + n$ reaction, $|M_{GT}/M_F|
= 1.26$ and thus $a \simeq -0.1$.  However, as we have shown
elsewhere~\cite{invbeta}, due to recoil
and weak magnetism corrections of order $1/M_p$, where $M_p$ is the
proton mass, the coefficient $a$ varies quite rapidly with neutrino
energy and changes sign near $E_{\nu}$ = 15 MeV, becoming positive at
higher energies (see also Ref.~\cite{Perkins}).
In fact, after averaging over the $\bar{\nu}_e$
spectrum, which is taken as before as Fermi-Dirac with a temperature
$T$ = 5 MeV, we obtain $\langle a \rangle \simeq +0.08$.  

As stated above, for $\bar{\nu}_e + p \rightarrow e^+ + n$ one expects
$N \simeq 10^4$ events in SK.  This would imply $\delta (\cos\theta)
\simeq 0.2$ if $a = -0.1$ and a similar error if $a = +0.08$ and the
temperature is {\it known} to be $T$ = 5 MeV (otherwise an uncertainty
in the temperature would obviously cause an additional uncertainty in
$\cos\theta$).  The temperature can be measured from the shape of the
positron spectrum, with precision of order
1\%~\cite{SKpaper,SNOpaper}.  Thus, even though the asymmetry
parameter $a$ is quite small, the number of events is large enough
that this technique in SK could give a reasonable pointing error.
There are also events of this type in the light water of SNO.
However, $N \simeq 400$, so this would imply $\delta (\cos\theta)
\simeq 1.0$, which is too large to be useful.

The reactions on deuterons in the $(M_p \rightarrow \infty)$
approximation are pure Gamow-Teller and thus $a = -1/3$.  We will, for
the sake of an estimate, assume that for the reactions on deuterons
the energy dependence of $a$ can be neglected (though see
Ref.~\cite{Fayans}).  We assume optimistically that the signal in SNO
of the reactions $\bar{\nu}_e + d \rightarrow e^+ + n + n$ and $\nu_e
+ d \rightarrow e^- + p + p$, can be combined, so $N \simeq 160$.
With $a = - 1/3$ this gives $\delta (\cos\theta) \simeq 0.5$, which is
again rather large.

In the reaction $\bar{\nu}_e + p \rightarrow e^+ + n$, the kinematics
dictate that the the outgoing neutrons have a forward angular
distribution.  If the positions of the positrons and the neutrons can
be separately determined, the vector between these points can give
some information on the neutrino direction, at least on a statistical
basis~\cite{Bemporad91}.  In fact, this effect was observed in the
Goesgen~\cite{Zacek}, and Chooz~\cite{Bemporad98} reactor experiments.
It is not currently possible to detect neutrons in SK or the light
water in SNO, and we do not consider this further.  However, this
technique may allow a scintillator detector to have some pointing
ability~\cite{invbeta}.

The techniques of this subsection will not allow the supernova to be
located anywhere near as precisely as by neutrino-electron scattering.
Nevertheless, they may provide an independent confirmation of the
neutrino-electron scattering results, which would increase the
confidence that a real supernova was seen.  For example, consider the
positrons from from $\bar{\nu}_e + p \rightarrow e^+ + n$ in SK.
Along the axis determined by neutrino-electron scattering, the
measured value should be $A_{FB} \simeq +0.04$, with
an error of 0.01, a four-sigma effect.  This can probably be improved
somewhat by considering only the highest-energy positrons.  While the
number of events will be reduced, the average $a$ will be
increased~\cite{invbeta}, thus improving the pointing ability.


\section{Triangulation}

For two detectors separated by a distance $d$, there will be a delay
between the arrival times of the neutrino pulse.  The magnitude of the
delay $\Delta t$ depends upon the angle $\theta$ between the supernova
direction and the axis connecting the two detectors.  Given a measured
time delay $\Delta t$, the unknown angle $\theta$ can be determined:
\begin{equation}
\cos\theta = \frac{\Delta t}{d}\,.
\end{equation}
The Earth diameter is $d \approx 40$ ms.  For a typical pair of
detectors, the time will be somewhat less; for SK and SNO, $d \approx
30$ ms.  The error on the time delay will cause an error on the
determination of $\theta$:
\begin{equation}
\delta(\cos\theta) = \frac{\delta(\Delta t)}{d}\,.
\end{equation}
Thus two detectors define a cone along their axis with opening
$\cos\theta$ and thickness $2 \times \delta(\cos\theta)$ in which the
supernova can lie.  Obviously, in order to have a reasonable pointing
accuracy from triangulation, one will need $\delta(\Delta t) \ll d$.
In what follows we discuss whether an appropriate time delay can be
defined, and what its error would likely be.

Note that this simple error analysis would have to be modified near
$\Delta t = \pm d$, since we must have $|\cos\theta| \le 1$, but we
ignore this complication.  Also, for convenience we use $\cos\theta$
rather than $\theta$ itself.  Naively, $\delta\theta =
\delta(\cos\theta)/\sin\theta$.  This is indeed valid for moderate
angles, but has spurious singularities at $\theta = 0,\pi$.  In fact,
for small $\delta\theta$, one has $\delta\theta \simeq \sqrt{2
\delta(\cos\theta)}$ near $\theta = 0,\pi$.

For now we will consider just two detectors, SK and SNO, taking events
from all reactions in SK and the light water in SNO.  These are about
$10^4$ and 400 events, respectively, mostly $\bar{\nu}_e + p
\rightarrow e^+ + n$.  The effect of multiple detectors will be
discussed later.  Further, we make the following assumptions: that the
detectors have perfect efficiency at all energies, perfect time
resolution and synchronization, no dead time, and a negligible
time-independent background.  In practice, these should be reasonable
assumptions.  Therefore, we consider that SK and the light water in
SNO are ideal detectors, identical except for size.  The event rates
in the two detectors should then be identical, except for
normalization and fluctuations, and a delay $\Delta t$.  That is, we
are considering the best that triangulation could do under any
circumstances, limited only by statistical errors.

The supernova neutrino pulse and hence also the observable scattering
rate is assumed to have a short rise, followed by a relatively slow
decline.  The total duration of the risetime is unknown, but in many
supernova models it is of order 100 ms~\cite{SNmodels1}.  The total
duration of the decaying phase is much longer, and in most supernova
models is a few seconds~\cite{SNmodels1}.  However, the {\it observed}
duration of SN1987A was clearly longer, about 10 s, and we will base
our estimates on that.

If we wish to assume as little as possible about the form of the event
rate, we could simply consider the shift in the average arrival time
$\langle t \rangle$ between the two detectors.  The expected value is
just $\Delta t$.  The error on the determination of the average
arrival time is the width of the pulse divided by the square root of
the number of events.  Quite generally, this must be of order $3 {\rm\
s}/\sqrt{10^4} \simeq 30$ ms for SK and $3/\sqrt{400} \simeq 150$ ms
for SNO, and therefore not useful for triangulation.  One can show
that a Kolmogorov-Smirnov test for a delay between the SK and SNO data
leads to the same result for the error.

In our previous studies of the effect of a $\nu_\tau$ mass on the
neutral-current event rate in SK or SNO~\cite{SKpaper,SNOpaper}, we
compared the average arrival time from the neutral-current events to
the average arrival time from the charged-current events.  We showed
that this allows the detection of a delay and the extraction of a
$\nu_\tau$ mass with only minimal assumptions about the time
dependences of the event rates.  The results obtained demonstrated the
best that one could do without assuming specific models for the event
rates.

Evidently, the triangulation problem will require more detailed
assumptions about the time dependence of the event rate.  As noted
above, here we have an additional piece of information, that besides
the delay $\Delta t$, the event rates in SK and the light water in SNO
should differ only in normalization and fluctuations.  As a schematic
model, we consider an event rate which consists of an exponential rise
with a short time scale ($\tau_1 \simeq 30$ ms), followed by slower
exponential decay ($\tau_2 \simeq 3$ s).  The point of transition
between the two is labeled $t_0$.  In Fig.~2, the normalized event
rate $f(t)$ is shown.  We show below how the results depend upon the
timescales $\tau_1$ and $\tau_2$.  Since SK will have many more events
than SNO, one could effectively measure the event rate in SK and use
that to replace our assumed form.


\subsection{Zero risetime case}

We consider first an even simpler model, in which the risetime $\tau_1
= 0$.  Then the normalized event rate can be written
\begin{equation}
f(t) = \frac{1}{\tau_2} \exp\left[-\frac{(t - t_0)}{\tau_2}\right],
\phantom{xxx} t > t_0\,,
\end{equation}
and is zero otherwise.  If $f(t)$ is guaranteed to have an infinitely
sharp edge, as above, one can show (see
Refs.~\cite{Hogg,Dalenius,Eadie}) that the best estimator (i.e., the
technique with the smallest error) of the edge $t_0$ is
\begin{equation}
t_0 \simeq t_1 - \frac{\tau_2}{N}\,,
\end{equation}
where $t_1$ is the measured time of the first event.  The error on the
determination of $t_0$ is
\begin{equation}
\delta(t_0) = \frac{\tau_2}{N}\,.
\end{equation}
Here $\tau_2/N$ is simply the spacing between events near the peak.
For a more general $f(t)$, but still with a sharp edge at $t_0$, one
would simply replace $\tau_2$ by $1/f(t_0)$.  In fact, the shape of
$f(t)$ is irrelevant except for its effect on the peak rate, i.e.,
$f(t_0)$.  So long as $f(t)$ has a sharp edge and the right total
duration, allowing a more general time dependence would therefore not
change the results significantly.  That is, only the long timescale
$\tau_2$ is important, since it determines the event spacing near
$t_0$.

The event rate in SNO will consist of $N$ events sampled from
$f(t,t_0)$, and the event rate in SK will consist of $N'$ events
sampled from $f(t,t_0')$ (where here we show the offset explicitly).
The parameters $t_0$ and $t_0'$ can be extracted from the times of the
first events as above.  Then $\Delta t = t_0 - t_0'$, and its error
will be dominated by the error on $t_0$ as extracted at SNO, so that
$\delta (\Delta t) \simeq \delta t_0$.  For $\tau_2 = 3$ s, this
idealized model would allow the offset of the edge to be measured to
$\simeq 0.3$ ms in SK, and $\simeq 8$ ms in SNO.  This gives
$\delta(\cos\theta) \simeq 0.25$ at one sigma.

This technique of using the first event only works if there is no
time-independent background and if there is absolutely no tail of
$f(t)$ before $t_0$.  In either case, the fluctuation of a single
unwanted event can change the extracted delay in an unpredictable way.
At the cost of an increase in the error, this technique could be made
robust by looking at the average time of the first few events.  In all
of the other techniques discussed in this paper, the role of the
time-independent background is negligible.


\begin{figure}[t]
\epsfxsize=3.25in \epsfbox{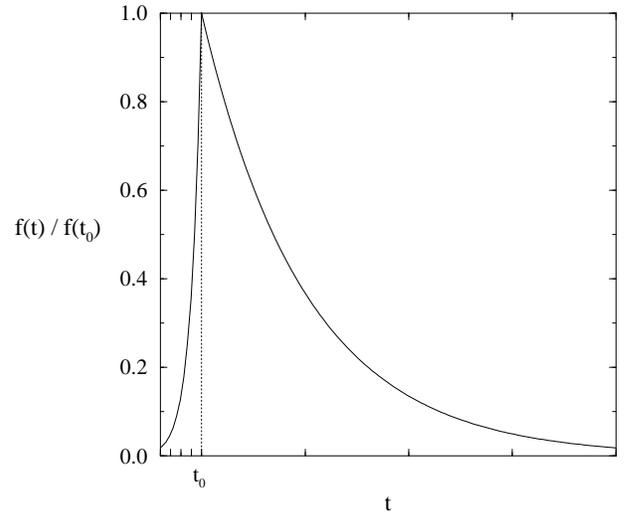}
\caption{The schematic form of the normalized event rate $f(t)$ is
shown.  To the left of $t_0$ there is an exponential rise with time
constant $\tau_1$.  To the right of $t_0$ there is an exponential
decay with time constant $\tau_2$.  The tick marks on the $t$-axis are
in units of the respective time constants $\tau_1$ and $\tau_2$.  For
clarity of display, we used $\tau_1/\tau_2 \simeq 10^{-1}$ in the
figure, instead of the $\tau_1/\tau_2 \simeq 10^{-2}$ assumed in the
analysis.}
\end{figure}

\subsection{Nonzero risetime case}

The above model has somewhat limited use, since the assumption of a
zero risetime does not seem to be justified.  As noted, the supernova
models suggest a nonzero risetime, of order $\tau_1 \simeq 30$ ms,
related to the shock propagation time across the supernova core.  If
the risetime is nonzero, the results of the previous subsection cannot
be used.  The error on $t_0$ is only given by the spacing between
events at the peak if the edge is sharp.  In this subsection, we allow
a nonzero risetime and show that the results have a qualitatively
different dependence on the parameters than in the previous case.  In
addition, softening the leading edge will obviously make the
triangulation error larger.

For the normalized event rate, we take
\begin{eqnarray}
f(t) & = &
\alpha_1 \times \frac{1}{\tau_1}
\exp\left[+\frac{(t - t_0)}{\tau_1}\right],
\phantom{xxx} t < t_0
\\
f(t) & = &
\alpha_2 \times \frac{1}{\tau_2}
\exp\left[-\frac{(t - t_0)}{\tau_2}\right],
\phantom{xxx} t > t_0\,.
\end{eqnarray}
where
\begin{equation}
\alpha_1 = \frac{\tau_1}{\tau_1 + \tau_2},
\phantom{xxxxxx}
\alpha_2 = \frac{\tau_2}{\tau_1 + \tau_2}\,.
\end{equation}
Then $f(t)$ is a normalized probability density function built out of
two exponentials, and joined continuously at $t = t_0$.  In what
follows, we assume that this form of $f(t)$ is known to be correct and
that $\tau_1$ and $\tau_2$ are {\it known}.

As above, the event rate in SNO will consist of $N$ events sampled
from $f(t,t_0)$, and the event rate in SK will consist of $N'$ events
sampled from $f(t,t_0')$.  Then $\Delta t = t_0 - t_0'$, and again
$\delta (\Delta t) \simeq \delta t_0$, since the SNO error dominates.
We consider only the statistical error determined by the number of
counts.  Any uncertainties in the form of $f(t)$ or its parameters
will only increase the error.  As noted, we want to determine the {\it
minimal} error on the triangulation.

This model, while simple, contains the essential timescales and an
adjustable offset.  More general models for the event rate must
reproduce these timescales in order to be physically plausible, and so
the final value for the error would be close to what is obtained here.
To define the time delay between the two pulses, we have used the
offset at which the rising and falling exponentials are joined.
However, in the final result for the error on the delay, the
particular way in which the offset time is defined drops out and the
result is therefore quite general.  That is, this simple model for the
event rate is general enough for calculating the statistical error on
the measured delay.

These considerations lead to a well-posed statistical problem: If $N$
events are sampled from a {\it known} distribution $f(t,t_0)$, how
well can $t_0$ be determined?  The Rao-Cramer
theorem~\cite{Hogg,Kendall} provides an answer to this question.  This
theorem allows one to calculate the minimum possible variance on the
determination of a parameter (here $t_0$), by any technique
whatsoever.  This minimum variance can be achieved when all of the
data is used as ``efficiently'' as possible, which is frequently
possible in practice.  One requirement of the theorem is that the
domain of positive probability must be independent of the parameter to
be determined.  This condition is obviously not met for a zero
risetime, since then the domain is $(t_0,\infty)$.  For a nonzero
risetime, the domain is technically $(-\infty,\infty)$, independent of
$t_0$, and so the theorem applies.  The minimum possible variance on
the determination of $t_0$ is:
\begin{equation}
\frac{1}{\left( \delta t_0 \right)^2_{\rm min}} = 
N \times
\int dt \, f(t,t_0)
\left[\frac{\partial \ln f(t,t_0)}{\partial t_0}\right]^2\,.
\end{equation}
This is the general form for an arbitrary parameter $t_0$.  When $t_0$
is a translation parameter, i.e., $f(t,t_0)$ depends only on $t -
t_0$, this reduces to
\begin{eqnarray}
\frac{1}{\left( \delta t_0 \right)^2_{\rm min}} & = & 
N \times
\int dt \, f(t,t_0)
\left[\frac{\partial \ln f(t,t_0)}{\partial t}\right]^2 \\
& = & 
N \times
\int dt \, 
\frac{\left[\partial f(t,t_0)/\partial t\right]^2}{f(t,t_0)}\,.
\end{eqnarray}
The latter form (the DuMond form) was independently derived by
another method in Ref.~\cite{DuMond}.  For the particular choice of
$f(t,t_0)$ above, this reduces to
\begin{equation}
\frac{1}{\left( \delta t_0 \right)^2_{\rm min}} = 
N \times
\left(\alpha_1/\tau_1^2 + \alpha_2/\tau_2^2\right)\,.
\end{equation}
For $\tau_1 \ll \tau_2$, the minimum error is then
\begin{equation}
\left(\delta t_0\right)_{\rm min} \simeq 
\frac{\sqrt{\tau_1 \tau_2}}{\sqrt{N}} \simeq
\frac{\tau_1}{\sqrt{N_1}}\,.
\end{equation}
Note that $N_1 \simeq N (\tau_1/\tau_2)$ is the the number of events
in the rising part of the pulse.  Since the rise is the sharpest
feature in $f(t)$, it is unsurprising that it contains almost all of
the information about $t_0$.  The total number of events $N$ is fixed
by the supernova binding energy release.  A change in the total
duration of the pulse, i.e., $\tau_2$, would therefore affect the peak
event rate and hence the fraction of events in the leading edge, i.e.,
$N_1/N = \alpha_1 \simeq \tau_1/\tau_2$.
That is, for $N$ fixed, we are considering how a change in the assumed
value of $\tau_2$ would affect the timing sensitivity; note that
$\tau_2$ appears only via the fraction of events in the leading part
of the pulse.
For a more general $f(t)$,
one would replace $\tau_1/\tau_2$ by this fraction computed directly.

For SNO, $N_1 \simeq 10^{-2} \times 400 \simeq 4$, so $\delta(t_0)
\simeq 30 {\rm\ ms}/\sqrt{4} \simeq 15$ ms.  Since SK has about 25
times more events, the corresponding error would be about 3 ms.
Therefore, the error on the delay is $\delta(\Delta t) \simeq 15$ ms
and $\delta(\cos\theta) \simeq 0.50$ at one sigma.  We have not yet
specified the method for extracting $t_0$ and hence $\Delta t$ from
the data.  That is exactly the point of the Rao-Cramer theorem -- that
one can determine the minimum possible error without having to try all
possible methods.  A possible technique which should come close to
achieving this minimal error is discussed below.

For the two cases, zero and nonzero risetime, we used different
mathematical techniques which were applicable only in one case or the
other.  This may seem like an artificial distinction, and that these
two cases do not naturally limit to each other.  In particular, it may
seem incompatible that in the first case the error $\sim 1/N$, while
in the second the error $\sim 1/\sqrt{N}$.  Further, one obviously
cannot take $\tau_1 \rightarrow 0$ in the results of this subsection.

However, there is physically a natural joining of the two results.
For the case of $\tau_1 = 0$, we found that $\delta(t_0) \simeq 8$ ms
for SNO.  For the case of $\tau_1 > 0$, we chose $\tau_1 = 30$ ms and
found $N_1 \simeq 4$ and $\delta(t_0) \simeq 15$ ms for SNO.  If we
reduce $\tau_1$ by a factor 4, so that $\tau_1 = 7.5$ ms, then this
error is reduced by a factor 2 so that $\delta(t_0) \simeq 7.5$ ms.
But now there is only $N_1 \simeq 1$ event in the rising part of the
pulse.  At this point, the difference between $\tau_1 = 0$ and $\tau_1
\le 7.5$ ms becomes difficult to distinguish, i.e., the edge appears
sharp.  That is, the two techniques give the same numerical result for
the error at the boundary between the two cases.


\subsection{What will the event rate really look like?}

We considered two simple models for what the event rates might look
like.  Those models were of course crude, yet they illustrate how the
different timescales affect the final results.  The short rise ($\sim
\tau_1$) at the beginning of the pulse is a prominent feature, and it
provides most of the timing information.  The long decay ($\sim
\tau_2$) is important principally through how it affects the
normalization, i.e., the number of events before or near the maximum.

The risetime, set by the timescale $\tau_1$, may be smaller than
suggested in Refs.~\cite{SNmodels1}.  For example, in some of the
models considered in Refs.~\cite{SNmodels2}, the duration of the rise
does appear to be shorter than considered here.  In some cases, the
$\bar{\nu}_e$ luminosity rises nearly instantaneously to a given
value, and then more slowly to a peak value which is several times
larger.  In this case, a zero-risetime analysis may be appropriate,
and the error on $t_0$ depends on the spacing between events.
However, one must not use the peak event rate, but rather the event
rate at the point where the rise is no longer instantaneous.  This
will give an error several times larger than the zero-risetime case
used above, where the instantaneous portion rose all the way to the
peak.

The neutrino pulse duration, set by the timescale $\tau_2$, may also
be smaller than assumed here, and our results can easily be scaled
appropriately.  Note that in the zero-risetime case, the error $\sim
\tau_2$, and in the nonzero-risetime case, the error is
$\sim\sqrt{\tau_2}$.  Thus in the latter case even if $\tau_2 = 1$ s,
the triangulation error would improve by only a factor $\sqrt{3}$, and
so $\delta(\cos\theta) \simeq 0.3$, still rather large.  However, one
has to keep in mind that our choice $\tau_2 = 3$ s was motivated by
the SN1987A observation that 25\% of the events arrived at least 5
seconds after the start of the pulse.  While the statistics were poor,
a timescale of $\tau_2 = 1$ s or smaller seems to be unlikely.  It may
also be that the decay of the neutrino pulse is characterized by two
timescales - a quick drop, followed by a slower decline, e.g., a
simplification of the gradually decreasing timescale considered in
Ref.~\cite{BurLat}.  Even so, if the duration of SN1987A is
reproduced, there would be little change in the error on the delay.

The event rate may also have a much more complex structure than
assumed here.  For example, there could be oscillations or other sharp
features which could be used to define a time from which to measure
the delay.  However, since the most prominent possible sharp feature,
a zero-risetime edge, is not enough for a successful triangulation,
any such features should be less significant.  Moreover, one should
not forget that the whole point of the supernova early alert network
is to use an {\it automated} analysis to determine the direction from
the data.  For the result to be available essentially immediately, the
analysis should assume as little as possible about the shape of the
pulse.  That was part of our motivation to consider a simple model
characterized only by timescales which are reasonably well-known.

Note also that we neglected the events in the prompt burst of electron
neutrinos.  The expected number of events is very small (for example,
Ref.~\cite{Sutaria} has $N \sim 1$ in SNO), although there is
considerable variation in the predictions for the number of events and
the duration of this burst.  It seems unlikely that the timing error
would be small enough to be useful, or that the predictions are robust
enough to allow an automated analysis.

In any case, the scenario for attempted triangulation could be the
following: the detector with the largest statistics, e.g., SK, could
be used to determine the fraction $N_1/N$ of events in the leading
edge and the duration of the risetime $\tau_1$ (both of these will
have some errors, unlike what we assumed in using the Rao-Cramer
theorem).  This fraction would then specify which of the SNO events
were to be considered as coming from the leading edge.  The delay
could then be determined by the time difference of the average arrival
times of leading edge events in SK and SNO.  The error would then be
$\simeq \tau_1/\sqrt{N_1}$, where $N_1$ is the number of events in SNO
coming from the rising edge.  Thus this technique might approximately
attain the Rao-Cramer lower bound on the error.  Alternatively, if SK
determines the zero-risetime model is applicable, the delay could be
extracted from the time difference of first events in each detector.
In this case, the error is again determined by SNO, and would be the
event spacing near the peak.


\subsection{Comparison to other work}

In Ref.~\cite{Vagins}, the triangulation error for a zero-risetime
pulse using SK and SNO was also considered, with a final result of 1.3
ms, to be compared with our result of 8 ms.  The difference is due to
different input parameters.  We assumed 400 events in the light water
of SNO, an exponential decay of the event rate with $\tau_2 = 3$ s,
and a distance of $D = 10$ kpc.  In Ref.~\cite{Vagins}, it was assumed
that all flavors and all reactions could be combined (we argue against
this below), for $10^3$ events in total (at a distance of 10 kpc),
with half in the first 1 s.  The greater number of events and the
shorter assumed duration of the pulse make the event rate at peak 500
s$^{-1}$ instead of 133 s$^{-1}$.  In addition, the final error was
scaled to a distance of 8 kpc.  After correcting for these
differences, the results are in agreement.  The same holds for the
results in Ref.~\cite{Burrows}, where again a peak rate of 500
s$^{-1}$ for SNO and a zero risetime were assumed.

In Ref.~\cite{Virtue}, a different technique was proposed, which does
not explicitly specify whether or not the risetime is nonzero.  The
proposed technique begins by constructing the cumulative distributions
(this function increases by a step of $1/N$ at each event time, and is
discrete but not binned) for SK and SNO.  At least for the light water
events, these functions should be the same up to fluctuations and a
possible delay.  The proposal is to make a simple low-order fit to the
beginning of each cumulative distribution and to extract the delay
from the difference of the intercepts.  A preliminary delay error of
order 5 ms was presented (for $D = 10$ kpc).  However, the fit to the
cumulative distribution function does not yet take into account the
fact that the errors on successive steps are highly correlated
(because most of the data at a given step are the data from the
previous step).  Taking this into account will increase the error.  In
any case, the error from this proposed method cannot be smaller than
that from the time of the first event (zero risetime case) or the
Rao-Cramer result (nonzero risetime case), whichever is appropriate.


\section{Conclusions and discussion}

The final uncertainties, calculated for a canonical supernova 10 kpc
away and with a total energy release of $3 \times 10^{53}$ ergs, are
summarized in Table I.  These are all one sigma errors, though larger
confidence regions may be necessary for making a search for a
supernova.

Neutrino-electron scattering has the best pointing precision.
Moreover, the calculated precision is largely independent of
assumptions about the supernova model.  In particular, it is totally
independent of the time dependence of the event rate.  The isotropic
background from other reactions degrades the precision somewhat, but
it is still the most precise technique.

The angular asymmetry of positrons from the $\bar{\nu}_e + p
\rightarrow e^+ + n$ reaction, even when combined with the high
statistics of SK, does not give a comparably small pointing error.  It
makes it possible, however, to check that the signal is indeed coming
from the right direction.  The angular distributions from the
charged-current deuteron reactions are even weaker.

Under realistic assumptions about the numbers of events and the
timescales, triangulation with SK and SNO appears to be very
difficult, if not impossible.  Other tests for the time delay can be
considered.  However, for either a zero risetime or a nonzero
risetime, we have shown in the previous section what the smallest
possible errors are.  For fixed inputs, there is simply no way to do
better.

But, so far in this paper, all of the concrete results were based on
using just two detectors, SK and SNO, and taking all events in the
light water (these are dominated by the charged-current signal
$\bar{\nu}_e + p \rightarrow e^+ + n$).  Can the pointing accuracy, in
particular for the triangulation technique, improve if other existing
or planned detectors are used?

First, we stress again that for the method to succeed the signals in
different detectors must differ only in the normalization and
fluctuations, and a possible delay.  For example, that precludes
including the neutral current events (dominated by $\nu_{\mu}$,
$\nu_{\tau}$, and their antiparticles) from the heavy water portion of
SNO.  That is because one cannot guarantee that the time dependence of
the scattering rate for these events is the same as for the events in
the light water.  In fact, at the crucial early times, the supernova
models suggest that there are differences among the flavors.  Since
the time dependences of the luminosities and temperatures are not
known to the needed high precision, these differences cannot be
corrected for.

There are several detectors, existing or under construction, that will
observe a few to several hundred $\bar{\nu}_e + p \rightarrow e^+ + n$
events.  In particular the existing MACRO and LVD detectors in Gran
Sasso, and Borexino and KamLAND detectors under construction will be
clearly able to ``see'' a Galactic supernova (see
Refs.~\cite{MACRO,LVD,Borexino,KamLAND}, respectively), and can
undoubtedly contribute to the false signal elimination in the planned
early alert network.  However, since these detectors are based on
scintillation rather than on \v{C}erenkov light, it is not {\it a
priori} clear that the basic requirement of the similarity of response
is indeed fulfilled.  And even if that difficulty could be overcome,
the numbers of events in those detectors will be comparable to what is
expected for SNO.  Thus our estimate of the uncertainty associated
with triangulation in the last line of Table I is valid for them also.
With three detectors, there is in principle an improvement in the
pointing from triangulation.  However, that improvement is minimal if
the pointing error from any two detectors is of order half of the sky.

The AMANDA detector (or its successor) can perhaps observe a supernova
via the $\bar{\nu}_e + p \rightarrow e^+ + n$ reaction in a very large
target volume~\cite{Jacobsen}.  However, the principle used for
supernova detection will be a fluctuation in the (large) background
rate.  Over an interval of a few seconds, the background events
dominate the signal events, $N_B \gg N_S$, but $N_S \gg \sqrt{N_B}$.
The actual supernova events can only be distinguished in a statistical
sense.  Under these circumstances, it will not be possible to map out
the event rate well enough to make a precise measurement of $t_0$ or
some other appropriate time.

What would it take to make triangulation viable?  As a simple example,
we consider SK and a hypothetical second detector, also with $\simeq
10^4$ events at 10 kpc, and a separation of 30 ms between the two.
The second detector might be very similar to SK, or it might be
primarily sensitive to neutral-current reactions (in the latter case,
we assume, despite the strong cautions above, that the charged-current
and neutral-current event rates can be directly compared for timing
purposes).  For the event rate assumed in our main analysis, each
detector would have a timing error of $\delta(t_0) \simeq 3$ ms, so
that the triangulation pointing error would be $\delta(\cos\theta)
\simeq 0.15$.  Results for the more general case of combining several
detectors with different timing errors have been given in
Refs.~\cite{nu98,Vagins}.

The pointing error from the angular distributions always scales with
$1/\sqrt{N}$.  Under the assumption that the event rate risetime is
nonzero~\cite{SNmodels1}, the triangulation pointing error also scales with
$1/\sqrt{N}$.  Since $N \sim 1/D^2$, all of the errors scale linearly
with the distance $D$.  The triangulation measurement may then become
feasible if the distance to the supernova is significantly less than
10 kpc.  However, all of the other techniques improve by the same
factor.
(However, if the risetime were vanishing~\cite{SNmodels2}, then the
triangulation error would scale as $1/N \sim D^2$; see the discussion
above.)

Thus our analysis shows that a Galactic supernova can indeed be
located by its neutrino signal, and that among the possible methods,
the best technique by a large margin is neutrino-electron scattering
in a water \v{C}erenkov detector.  Currently, either SK or SNO can
separately make this measurement.
In the above, we considered the directional information from the
neutrino data alone.  The operators of the alert network or the
astronomers themselves can of course combine these results with a
Galactic model of where a supernova is likely to be.
Our results indicating that triangulation will be very difficult
do not mean that the data sharing among
different detectors is not worthwhile.  Only a coincidence of two or
more detectors can eliminate false alarms, and be the basis of a
reliable early alert system.


\section*{ACKNOWLEDGMENTS}

This work was supported in part by the U.S. Department of Energy under
Grant No. DE-FG03-88ER-40397.  J.F.B. was supported by a Sherman
Fairchild fellowship from Caltech.  We thank Kate Scholberg, Alec
Habig, Mark Vagins, Adam Burrows, and the other participants in the
Supernova Early Alert Network Workshop for discussions on supernova
location, and Robert Sherman and Brad Filippone for discussions on
statistics.  In addition, we thank Mark Vagins for bringing this
problem to our attention.  We think that Jesse DuMond would have been
pleased that his Ref.~\cite{DuMond} was valuable to the person
(J.F.B.) who now occupies the office in which it was written forty-six
years earlier.



\newpage
\widetext

\begin{table}
\caption{One-sigma errors on how well the direction to the supernova
is defined by various techniques, at $D = 10$ kpc.  The other parameters
used are noted in the text.  For neutrino-electron scattering, the most
pessimistic background assumptions were used.}
\begin{tabular}{l|r}
technique & error \\
\hline
$\nu + e^-$ forward scattering (SK) &
 $\delta \theta \simeq 5^\circ$,
 $\delta(\cos\theta) \simeq 4 \times 10^{-3}$ \\
$\nu + e^-$ forward scattering (SNO) & 
 $\delta \theta \simeq 20^\circ$,
 $\delta(\cos\theta) \simeq 6 \times 10^{-2}$ \\
\hline
$\bar{\nu}_e + p$ angular distribution (SK) &
 $\delta(\cos\theta) \simeq 0.2$ \\
$\bar{\nu}_e + p$ angular distribution (SNO) &
 $\delta(\cos\theta) \simeq 1.0$ \\
$\nu_e + d, \bar{\nu}_e + d$ angular distributions (SNO) &
 $\delta(\cos\theta) \simeq 0.5$ \\
\hline
triangulation (SK and SNO) &
 $\delta(\cos\theta) \simeq 0.5$ \\
\end{tabular}
\end{table}

\narrowtext

\end{document}